\documentclass[prb,showpacs,twocolumn,preprintnumbers,amsmath,amssymb,floatfix]{revtex4-1}
\usepackage[dvips]{graphicx}
\usepackage[latin1]{inputenc}
\usepackage{stmaryrd}
\usepackage{wasysym}
\usepackage{subfigure}
\usepackage{booktabs}

\usepackage{color}
\usepackage{epstopdf}
\usepackage{pdfpages}

\begin{document}
\draft
\title{Vortex Lattice Melting of a NbSe$_2$ single grain probed by Ultrasensitive Cantilever Magnetometry}

\author{L. Bossoni,$^{1,2}$ P. Carretta,$^{1}$  M. Poggio$^{3}$}
\address{$^{1}$ Department of Physics, University of
Pavia-CNISM, Via Bassi 6, I-27100 Pavia, Italy}
\address{$^{2}$ Department of Physics "E. Amaldi", University of Roma Tre-CNISM, I-00146 Roma, Italy}
\address{$^{3}$ Department of Physics, University of Basel,
Klingelbergstrasse 82, CH-4056 Basel, Switzerland.}

\begin{abstract}
Using dynamic cantilever magnetometry, we study the vortex lattice and its corresponding melting transition in a micrometer-size crystallite of superconducting NbSe$_2$. Measurements of the cantilever resonance frequency as a function of magnetic field and temperature respond to the magnetization of the vortex-lattice. The cantilever dissipation depends on thermally activated vortex creep motion, whose pinning energy barrier is found to be in good agreement with transport measurements on bulk samples. This approach reveals the phase diagram of the crystallite, and is applicable to other micro- or nanometer-scale superconducting samples.
\end{abstract}

\maketitle

\draft

\narrowtext

\small
The study of vortex physics in type-II superconductors touches on several phenomena, including hydrodynamics, electromagnetism and quantum field theory. The interplay between thermal fluctuations, vortex repulsion/attraction and the role of quenched disorder contributes to create a complex and interesting scenario.\cite{Blatter1994,Jang2011} In fact,
a dynamical and structural transition from vortex solid/glass to vortex liquid is often observed, and is particularly manifest in layered superconductors, where the melting line usually appears well below the upper critical field $H_{c2}$. A study of this transition is appealing not only for its fundamental aspects, but also in the light of the practical limitations related to the occurrence of a liquid vortex phase, where the dissipationless state peculiar to superconductivity vanishes. The melting transition has been intensively investigated over the last few years, in a series of theoretical\cite{Brandt95,Clem1991,Glazman1991} and experimental studies.\cite{Hess1990,Recchia95,Keimer}
	
Among the most widely employed techniques to characterize the vortex lattice (VL), one finds Scanning Tunneling Microscopy (STM),\cite{Troyanovski1999} magnetic decoration,\cite{essmann} Scanning Electron Microscopy (SEM),\cite{Vinnikov} Magnetic Force Microscopy (MFM).\cite{moser} Some drawbacks of these methods are the sensitivity to the topography of the sample surface, and the applicability only at fields in the mT range. On the other hand resistivity, ac-susceptibility,\cite{Palstra} Nuclear Magnetic Resonance (NMR),\cite{Rigamonti1998,Delrieu, Kim1965,Carretta1993,Bossoni2012,Bossoni2013,Oh2011,Corti96} Muon Spin Rotation ($\mu$SR),\cite{Lee1993} and neutron scattering spectroscopy,\cite{Lynn} allow the application of stronger fields, but require large samples of at least a few mm$^3$ or cm$^3$. The combination of sub-millimetric samples, and fields in the Tesla range has not been often encountered.

Dynamic cantilever magnetometry\cite{Stipe2001} is able to fill this gap, as it allows the use of nm-$\mu$m size samples, and fields ranging from the mT up to the Tesla range. The high sensitivity of the technique allows for the detection of the weak magnetic response of micro- and nanometer-scale samples. This sensitivity and its continued improvement is a direct result of recent advances in the fabrication of ultrasensitive Si cantilevers,\cite{Poggiorev} as demonstrated by recent measurements of the persistent currents in normal metal rings,\cite{BJ2009} of the magnetization of superconducting nanostructures,\cite{Jang2011} and of magnetization reversal in a single iron-filled carbon nanotube\cite{Banerjee2010} and a single Ni nanorod.\cite{Lee2012}

In this letter, a micrometer-scale sample of a well-known type-II superconductor is investigated by cantilever magnetometry.
NbSe$_2$ is chosen as it is a layered s-wave superconductor, with $T_c\sim 7.2$ K.\cite{Leupold,Garoche,Banerjee1997} It is known to show multiband superconductivity, with distinct small and large superconducting gaps on different sheets of the Fermi surface.\cite{Boaknin2003,Fletcher2007} Furthermore the vortex phase of NbSe$_2$ is characterized by a plastic flow which
dominates the dynamics.\cite{Battacharia1993}
By monitoring the cantilever resonance frequency and dissipation, we measure the behavior of both the VL magnetization and the dynamical response of the flux lines lattice (FLL). In particular, the pinning energy barriers of the thermally activated creep motion are derived, and the mixed phase diagram of the material is drawn, for magnetic fields up to 6 Tesla.\\

A powder of superconducting NbSe$_2$ is first characterized by SQUID magnetometry and by SEM microscopy.\cite{READE} The static spin susceptibility shows $T_c(0)=7.2$ K, while the average grain size of the crystallites is 1.4 $\mu$m. A superconducting grain of volume $\sim$ 16 $\mu$m$^3$ is chosen with a glass needle using precision micromanipulators, combined with an optical microscope.
The grain is attached to the cantilever tip, with epoxy (Gatan G1).
\begin{figure}[h!]
\centering
\includegraphics[height=5.5cm, keepaspectratio]{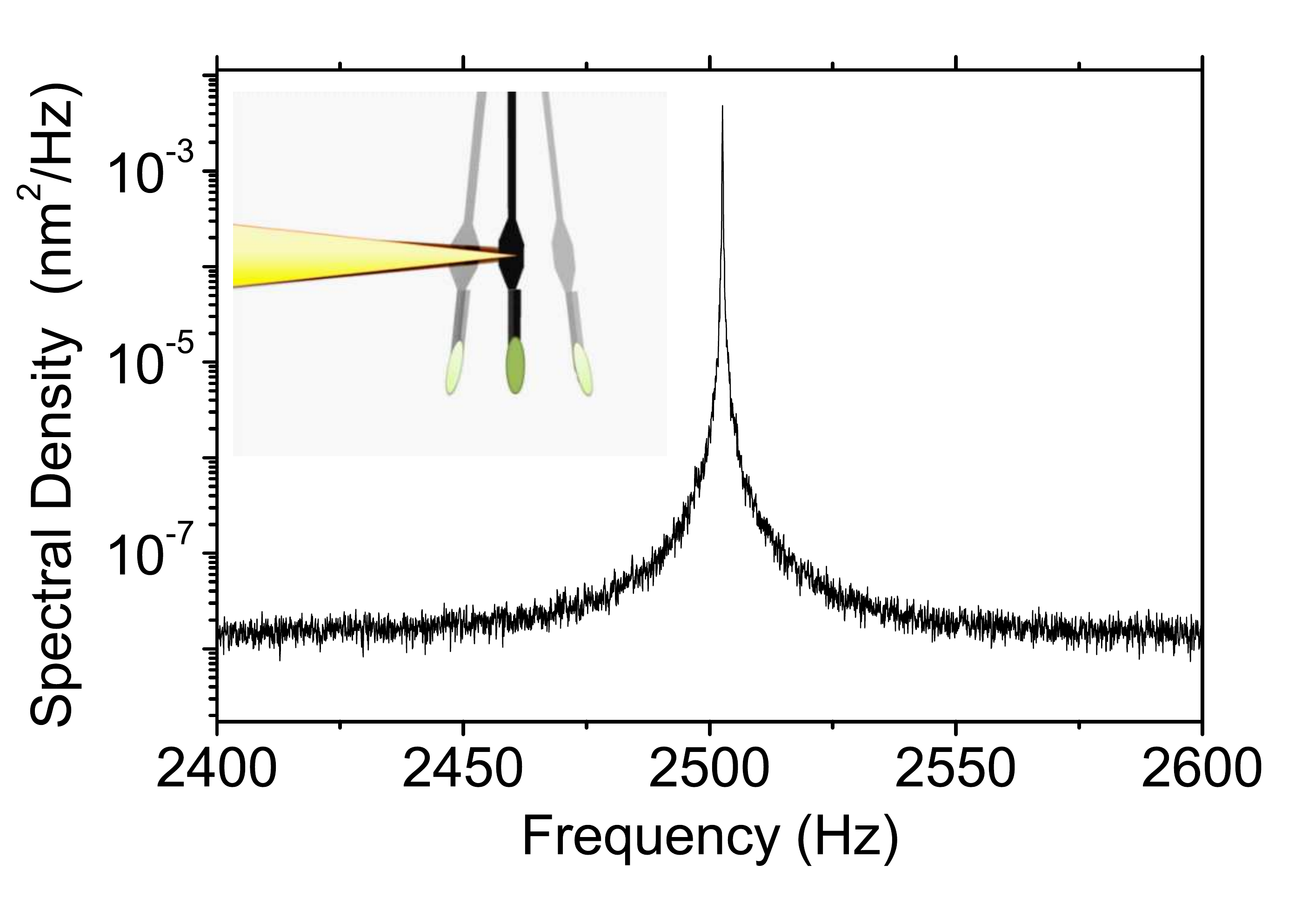}
\caption{Spectral density of the thermal motion of the cantilever's fundamental mode, measured at 4.3 K, and zero field. Inset: a sketch of the oscillating cantilever, and mounted sample.}
\label{sk}
\end{figure}
The single-crystal Si cantilever is 105 $\mu$m long, 4 $\mu$m wide, and 0.1 $\mu$m thick and includes a 18 $\mu$m long, 1 $\mu$m thick mass on its end. It has a small spring constant $k=80$ $\mu$N/m, with low intrinsic dissipation, that is ideal for detecting small forces.
The motion of the lever is detected using an optical fiber interferometer operating at 1550 nm with 20 nW of optical power incident onto a 12-$\mu m$-wide paddle (Fig. \ref{sk}). The sample and cantilever are inserted into an ultra high vacuum (UHV) chamber at the bottom of a $^3$He continuous-flow cryostat, mechanically insulated from the ground and equipped with a 6 T superconducting magnet, with the field applied along the cantilever axis.

The sample-mounted cantilever's resonance frequency $\nu_0 = \omega_0 / (2 \pi)$ and its mechanical dissipation $\Gamma$ are measured through the "ringdown" method, as described by Stipe \textit{et al.}\cite{Stipe2001} The cantilever is oscillated at its natural resonance frequency, with a root mean square amplitude of typically 10 to 20 nm, using a piezoelectric disk and a gain-controlled positive feedback loop. The drive circuit is then abruptly grounded and the cantilever oscillation amplitude decays until thermal equilibrium is recovered. The cantilever response is fit with an exponentially decaying sinusoid to extract the resonant frequency $\nu_0$ and the decay time constant $\tau$. The same results can be obtained by measuring the spectral density of the cantilever's thermal motion, as shown in Fig. \ref{sk}, and fitting the fundamental mode to a Lorentzian.\\

\begin{figure}[h!]
\centering
\includegraphics[height=6cm, keepaspectratio]{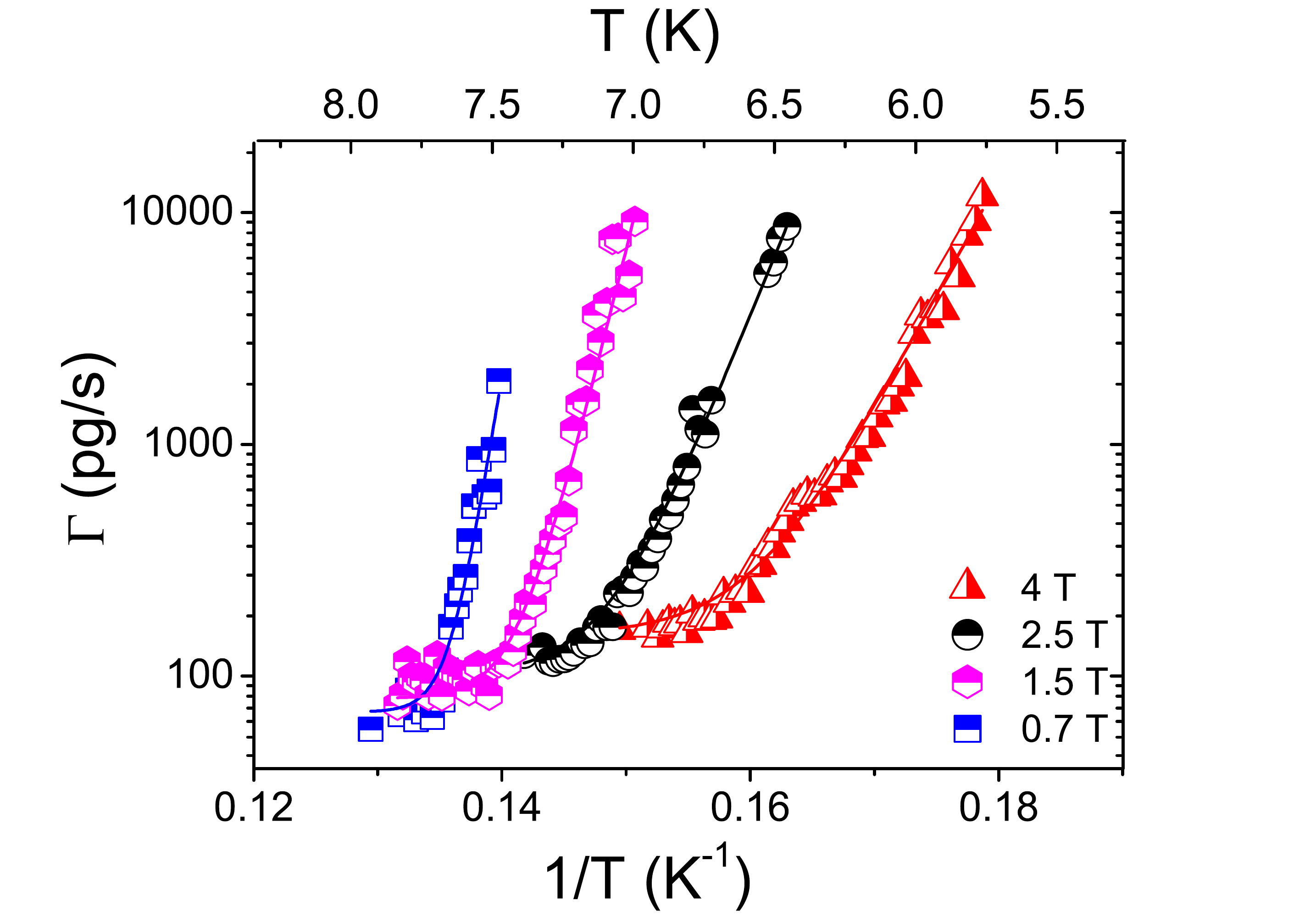}
\caption{Temperature evolution of the cantilever energy dissipation, measured after field-cooling, at different fields. The upturn of $\Gamma$ marks the onset of the superconductivity transition, as discussed in the text. The solid lines are the best fits to equation \ref{gamma}.}
\label{GammaVsT}
\end{figure}

\begin{figure}[h!]
\centering
\includegraphics[height=9cm, keepaspectratio]{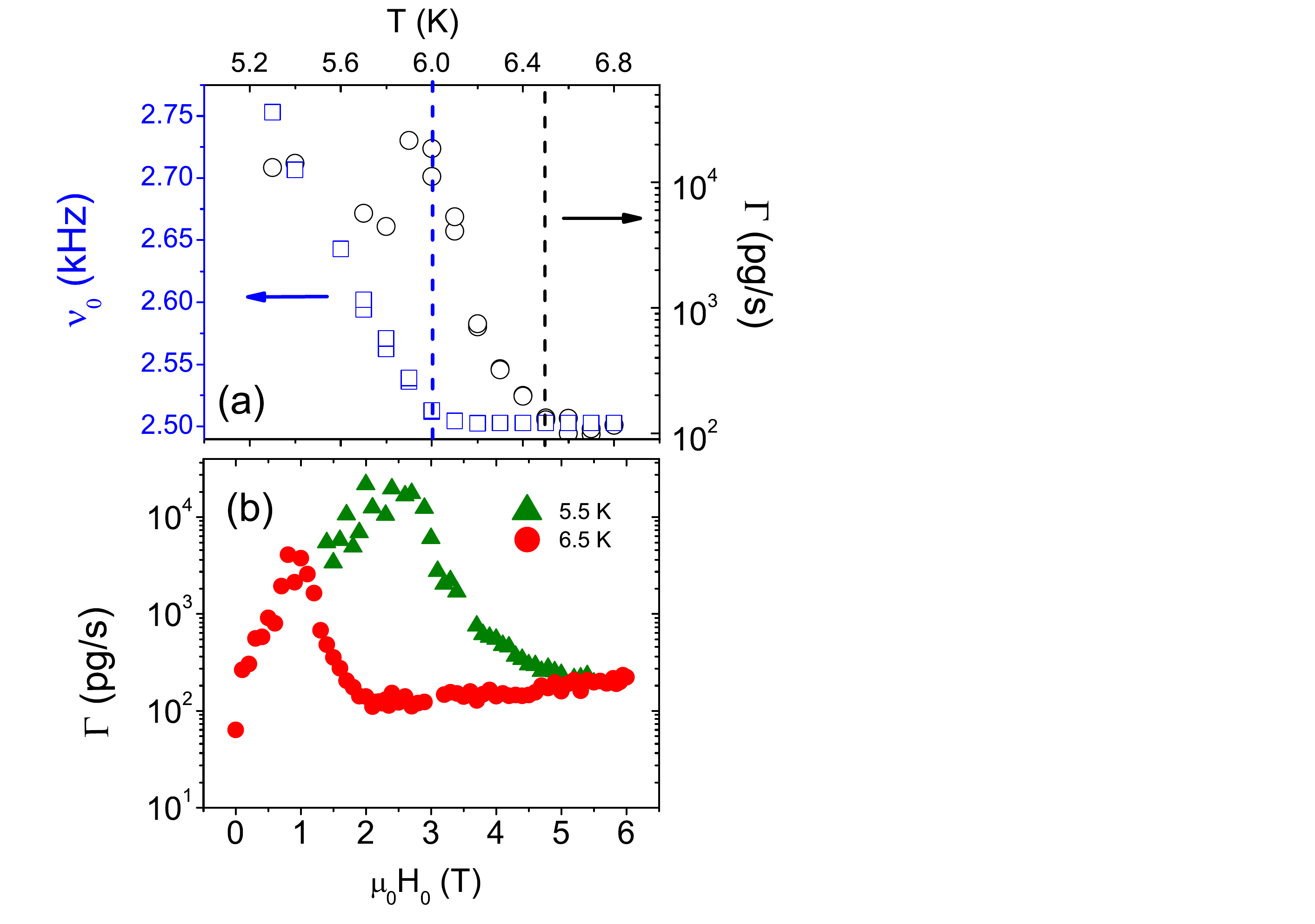}
\caption{\textbf{(a)} Temperature evolution of $\Gamma$ (open black circles) and fundamental mode frequency $\nu_0$ (open blue squares) of the sample-mounted cantilever, at 2 T. The increase of the cantilever natural frequency is found below the dissipation increase, at each field. The dotted lines mark the transition from normal phase to the liquid vortex phase (black), and then to the solid vortex phase (blue). \textbf{(b)} Cantilever dissipation measured at 6.5 K (red circles) and 5.5 K (green triangles), by sweeping the field from 6 T  to zero. Upon decreasing the magnetic field, the dissipation decreases, at the vortex freezing transition.}
\label{peak}
\end{figure}

The study of the cantilever mechanical response, as a function of the temperature, reveals a sudden increase in the energy dissipation $\Gamma$, close to $T_c$. Such an effect, reported in Fig. \ref{GammaVsT}, has been observed earlier in other superconductors,\cite{Gammel1988,Meyer2006,Kisiel2011} and it has been interpreted as the sudden change of the sample magnetization, due to the Meissner-Ochsenfel effect. Furthermore, in type-II superconductors, the \textit{incomplete} Meissner effect related to the FLL penetration, and the thermal fluctuations of the vortices competing with the pinning mechanism, can significantly affect the cantilever elastic response. In fact, the upturn in cantilever dissipation, observed at $T_c$, can only occur through a non-conservative energy relaxation mechanism. 
In this case the dominant dissipative mechanism is ascribed to the flux-creep motion of vortices, hopping among metastable energy minima, generated by the pinning potential, as it will be discussed subsequently.
Remarkably, at a temperature systematically \textit{below} the increase in $\Gamma$, an abrupt increase in the cantilever resonance frequency $\nu_0$ is observed, as shown in Fig. \ref{peak}(a).
Moreover, a study of $\Gamma$ as a function of $H_0$ reveals a peak, denoting a phase transition, in the vortex matter, as shown in Fig. \ref{peak}(b).\\

\begin{figure}[h!]
\centering
\includegraphics[height=6.8cm, keepaspectratio]{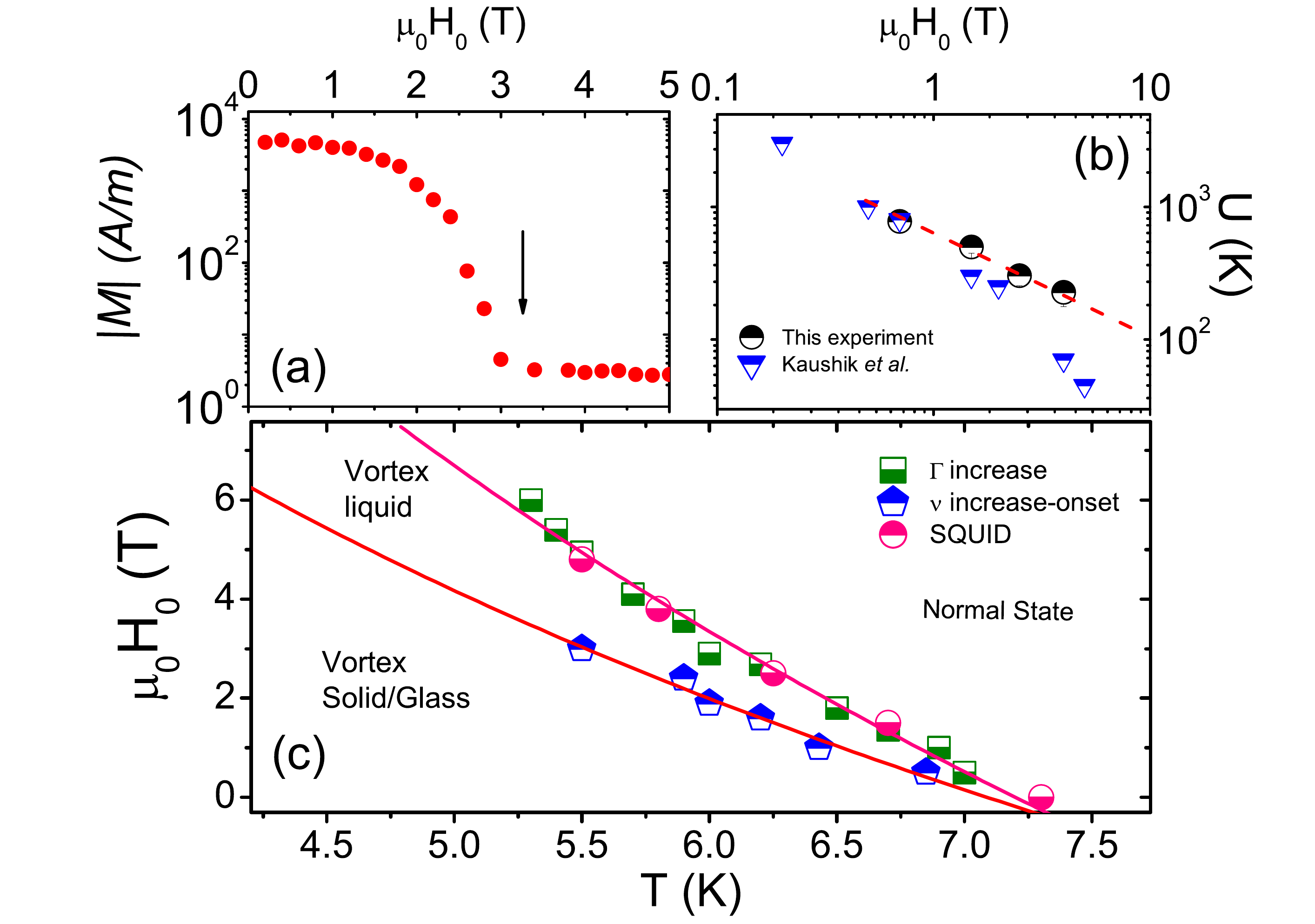}
\caption{\textbf{(a)} The absolute value of the superconducting grain magnetization is plotted, as a function of the magnetic field at 6 K, in agreement with Eq. (6). The arrow marks $H_{c2}$. \textbf{(b)} Energy barrier of the pinning as a function of the field, for the NbSe$_2$ grain (black circles), as compared to the result presented in Ref. \onlinecite{Kaushik}, on the same compound (blue triangles). The red dashed line is a guide for the eye. \textbf{(c)} The phase diagram of the NbSe$_2$ grain: the $H_{c2}$ line is derived from the $\Gamma$ upturn (green squares) and the SQUID measurement. The frequency increase (red triangles) marks the melting transition, as discussed in the text. }
\label{diagram}
\end{figure}

An analytic expression of the magnetization of the sample can be written as a function of the cantilever's resonance frequency $\nu_0$ and its mechanical dissipation $\Gamma$. Since the relaxation of the whole VL magnetization takes place over a timescale much longer than $1/\nu_0$ (typically few hours) the total energy of the sample-mounted cantilever, in the superconducting region, can be written as\cite{nota}
\begin{equation}
\mathcal{E}=\frac{1}{2}k(l_c\theta)^2-V \mu_0\mathcal{M}\cdot \mathbf{H_0} \label{energia}
\end{equation}
where $l_c$ is the cantilever length, $k$ the cantilever elastic constant, $\mathbf{H_0}$ the external magnetic field, $V$ the sample volume and $\mathcal{M}$ the grain magnetization.  \\
The scalar product in Eq.(\ref{energia}) gives a $\cos \theta$ factor, which can be approximated up to the second order, for $\theta \ll 1$. Here the angle $\theta$ is formed between the vortex direction, which is fixed to the sample, and the direction of the applied field $\mathbf{H_0}$.
The angular dependence of the energy gives rise to a torque,
\begin{equation}
\tau=-\frac{\partial \mathcal{E}}{\partial \theta}=-(k l_c^2+ V \mathcal{M} \mu_0 H_0)\theta.
\end{equation}
Recalling the equation of motion for the damped harmonic oscillator,\cite{Weber2012} one can write:
\begin{equation}
m l_c^2 \frac{\partial^2 \theta(t)}{\partial t^2}+\Gamma l_c^2 \frac{\partial \theta(t)}{\partial t}+(k l_c^2+ V \mathcal{M} \mu_0 H_0)\theta(t)=0,
\end{equation}
The partial derivative equation has the following solution:\cite{nota}
\begin{equation}
\theta(t)=c e^{-t/\tau} \sin( \omega_0 t), \label{theta}
\end{equation}
The system indeed oscillates as an underdamped harmonic oscillator, where the frequency is given by
\begin{equation}
\omega_0=\sqrt{\frac{k}{m}+\frac{V \mathcal{M} \mu_0 H_0}{m l_c^2} - \frac{\Gamma^2}{4m^2}}.\label{NH}
\end{equation}
Since equation (\ref{NH}) shows a one-to-one correspondence between the sample magnetization and the measurable parameters ($\Gamma, \omega$), the magnetization can be expressed as 
\begin{equation}
\mathcal{M}\simeq \frac{m l_c^2}{V \mu_0 H}\left(  \omega^2-\frac{k}{m}\right),
\end{equation}
where the dissipation term has been neglected, because its value is negligible compared to the other two terms. The absolute value of $\mathcal{M}$, obtained at 6 K, is shown in Fig. \ref{diagram}(a), where the onset of the superconducting transition is marked by an arrow.\\

As far as the temperature dependence of the dissipation is concerned, $\Gamma(T,H)$ is the sum of $\Gamma_c+\Gamma_v (T,H)$, i.e. the intrinsic cantilever losses plus the vortex loss.
By drawing an analogy between $\Gamma$, and the imaginary part of the ac susceptibility, or the magnetoresistivity, which are all strictly related to the energy dissipation induced by flux creep motion,\cite{Anderson} the data can be fit to the expression:\\
\begin{equation}
\Gamma(T,H)=\Gamma_c+\Gamma_0 e^{U(H)/T} \label{gamma}
\end{equation}
where $U$ represents the pinning energy barrier (in Kelvin) of the thermally activated vortex motion. Fig. \ref{GammaVsT} shows that the thermally activated model fits the experimental data, supporting the initial assumption. From the fit, the intrinsic cantilever dissipation turns out to be $\Gamma_c\simeq 80$ pg/s, while $\Gamma_0\simeq 10^{-20}$ pg/s.
Moreover a study of the pinning energy barrier $U$ as a function of the magnetic field intensity is reported in Fig. \ref{diagram}(b). At first one notices a power-law behavior (red dotted line), as expected for a vortex bundle motion.\cite{Tinkham} Indeed, a STM study on the same compound, although in powder form, shows the occurrence of a collective vortex bundle creep, taking place at 0.6 T, under the application of a strong current $J=0.4 J_c$.\cite{Troyanovski1999} In the same panel, the pinning barriers are compared to Ref.\onlinecite{Kaushik}, reporting a magnetoresistivity study on NbSe$_2$. The slight disagreement between the two data sets can be ascribed either to a powder effect, which in the transport measurements averages the activation barrier along the crystallographic directions, or to a small underestimation of $\Gamma_c$. \\

Just as the cantilever dissipation increases as the sample enters the vortex liquid phase from the normal phase, the cantilever resonance frequency increases as the sample makes the transition to the solid vortex phase. This increase reflects the stiffening effective cantilever spring constant due to the magnetization of the fixed vortex lines. In addition, as the vortices solidify, their hopping correlation time becomes long with respect to $\nu_0$.  As a result, the cantilever's low dissipation state should be restored along with the stiffening of its spring constant. As expected, Fig. 3(a) shows a rise in $\nu_0$ occurring at a lower temperature than the onset of the high dissipation state.  The expected simultaneous reduction in dissipation is partially obscured by the noisiness of the dissipation data in this temperature range.
Moreover, when $\Gamma$ is plotted as a function of the magnetic field, a peak is observed and ascribed to the crossover from the liquid to the solid vortex phase (Fig. \ref{peak}(b)). One may argue that such decrease of $\Gamma$ at low field is not related to the freezing transition, but is rather due to the diminished interaction of the field with the FLL. However, as the temperature decreases, the peak moves towards higher fields, thus ruling out the former hypothesis.
An analogous phenomenology was found by Gammel \textit{et al.},\cite{Gammel1988} using mechanical measurements on high temperature superconductors single crystals, with a surface of about 1 mm$^2$, 0.1 mm thick, and containing many twins.
However note that here the sample dimension is pushed to the limit of few $\mu m$, and a wider field range is explored. \\

Finally, a phase diagram of the mixed state of the NbSe$_2$ particle is drawn (Fig. \ref{diagram}(c)). The $\Gamma$ onset overlaps with the $H_{c2}$ data, measured on the powders by a SQUID magnetometer. 
The diagram allows the identification of the vortex liquid phase and the transition to the solid phase.

In conclusion, the present letter shows a cantilever magnetometry experiment, on a micrometer-sized NbSe$_2$.
The solution of the equation of motion results in an analytic expression for the vortex state magnetization, which depends on the measured parameters. 
The temperature and field dependence of the cantilever energy dissipation and oscillation frequency reveal the energy barrier of the pinning mechanism, as a function of the field.
Such results show that the ultrasensitive cantilever magnetometry is an effective technique for measuring the properties of VL in micro- and nanometer-scale samples, and that its results are directly comparable with macroscopic techniques. \\

Dr. C. Milanese and Dr. M. Mazzani are kindly acknowledged for the SEM and SQUID measurements.  Dr. M. Montinaro, F. Xue and A. Buchter are gratefully thanked for their technical assistance. P.C. and L.B. acknowledge financial support from Fondazione Cariplo (Research Grant No. 2011-0266).
M. P. acknowledges support from the Canton Aargau and the Swiss Nanoscience Institute.



\begin{references}
\bibitem{nota} Note that above the lower critical field, $H_{c1}$, the magnetization arises from the partial screening of the magnetic field by the VL. Hence electron paramagnetic contributions can be neglected with respect to that diamagnetic term.
\end{references}
\end{document}